\def\papertitle{Probing Low-Level Acoustic Attribute Encoding in CLAP Audio Embeddings}
\def\paperauthorA{Héctor Martel}
\def\paperauthorB{Joe Hennessy-Priest}
\def\paperauthorC{Taemin Cho}

\documentclass[twoside,a4paper]{article}
\usepackage{etoolbox}

\usepackage[print]{dafx26v2}

\usepackage{amsmath,amssymb,amsfonts,amsthm}
\usepackage{siunitx}
\usepackage{euscript}
\usepackage[T1]{fontenc}
\usepackage[utf8]{inputenc}
\usepackage{ifpdf}
\usepackage[english]{babel}
\usepackage{caption}
\usepackage{subcaption}
\usepackage[dvipsnames]{xcolor}
\usepackage{booktabs}
\usepackage{lipsum}
\usepackage{multirow}
\usepackage{makecell}

\input glyphtounicode
\ninept

\newcounter{numauth}
\newcounter{listcnt}
\newcommand\authcnt[1]{\ifdefined#1 \stepcounter{numauth} \fi}

\newcommand\addauth[1]{
\ifdefined#1 
\stepcounter{listcnt}
\ifnum \value{listcnt}<\value{numauth}
\appto\authorslist{, #1}
\else
\appto\authorslist{~and~#1}
\fi
\fi}
\authcnt{\paperauthorB}
\authcnt{\paperauthorC}
\authcnt{\paperauthorD}
\authcnt{\paperauthorE}
\authcnt{\paperauthorF}
\authcnt{\paperauthorG}
\authcnt{\paperauthorH}
\authcnt{\paperauthorI}
\authcnt{\paperauthorJ}
\def\authorslist{\paperauthorA}
\addauth{\paperauthorB}
\addauth{\paperauthorC}
\addauth{\paperauthorD}
\addauth{\paperauthorE}
\addauth{\paperauthorF}
\addauth{\paperauthorG}
\addauth{\paperauthorH}
\addauth{\paperauthorI}
\addauth{\paperauthorJ}

\usepackage{times}

\newif\ifpdf
\ifx\pdfoutput\relax
\else
   \ifcase\pdfoutput
      \pdffalse
   \else
      \pdftrue
   \fi
\fi

\ifpdf %
  \usepackage[pdftex,
    pdftitle={\papertitle},
    pdfauthor={\authorslist},
    pdfsubject={Proceedings of the 29th International Conference on Digital Audio Effects (DAFx26)},
    colorlinks=false, %
    bookmarksnumbered, %
    pdfstartview=XYZ %
  ]{hyperref}
  \usepackage[pdftex]{graphicx}
\else %
  \usepackage[dvips]{epsfig,graphicx}
  \usepackage[dvips,
    pdftitle={\papertitle},
    pdfauthor={\authorslist},
    pdfsubject={Proceedings of the 29th International Conference on Digital Audio Effects (DAFx26)},
    colorlinks=false, %
    bookmarksnumbered, %
    pdfstartview=XYZ %
  ]{hyperref}
\fi
\usepackage[hypcap=true]{caption}
\title{\papertitle}

\affiliation
{\authorslist %
}
{\href{https://bandlab.com}{BandLab Technologies, Singapore}\\
{\tt \href{mailto:hector.martel@bandlab.com}{hector.martel@bandlab.com}}
}

\begin{document}
\ifpdf %
  \DeclareGraphicsExtensions{.png,.jpg,.pdf}
\else  %
  \DeclareGraphicsExtensions{.eps}
\fi

\maketitle

\begin{abstract}
Audio foundation models are widely adopted as general-purpose feature extractors, yet the internal structure of their learned representations remains insufficiently understood.
In this work, we analyze CLAP audio embeddings through a probing framework, studying the encoding of three fundamental perceptual dimensions: reverberation (RT60), loudness (LUFS), and spectral content, measured via spectral centroid (SC) and relative pitch (RP).
Probes of increasing complexity are trained to predict each attribute from frozen embeddings across five datasets spanning noise, speech, monophonic musical notes, and music mixtures.
Our primary finding is that all of these attributes are reliably recoverable from the CLAP embedding space across the examined datasets.
Within this global picture, two encoding regimes emerge: RT60, LUFS, and RP are approximately linearly encoded, while SC requires non-linear probes.
Both regimes generalize across eight additional audio foundation models, with the notable exception that amplitude-invariant architectures discard loudness entirely by construction.
The identified linear feature directions are geometrically consistent across datasets for RT60 and LUFS, while highly domain-specific for RP.
Finally, we provide a qualitative demonstration of cross-modal consistency, showing that text embeddings of acoustic descriptors align geometrically with the identified RT60 feature direction.
\end{abstract}

\section{Introduction}
\label{sec:intro}

CLAP \cite{elizalde2022clap, elizalde2023clap} is a prominent audio-language foundation model that leverages contrastive learning to align audio and text in a shared embedding space. 
Trained on large collections of audio-text pairs, CLAP learns embeddings that capture both acoustic characteristics and semantic content, enabling applications such as audio retrieval, captioning \cite{zhang2025transformationaudioembeddingsinterpretable,li2025drcap}, audio quality assessment \cite{deshmukh2024pam,chung2025listen}, effect parameter estimation \cite{text2fx}, room impulse response synthesis \cite{vosoughi2025promptreverb}, and conditional audio generation \cite{kim2025tokensynth,yang2025flowsynth,pmlr-v202-huang23i}. 
The model has since been extended to incorporate temporal \cite{yuan2024t} and stereo \cite{seki2025spatial} information.

Despite this widespread adoption, the internal representation structure of CLAP remains insufficiently understood.
Probing studies of audio-language models have largely examined specific domains or model classes: early work probed speech embeddings for speaker and phonetic information \cite{Raj_2019}, while more recent analyses of Large Audio-Language Models (LALMs), a model class that does not encompass CLAP, probe high-level perceptual attributes such as language, gender, and emotion of speech \cite{yang2025audiolenscloserlookauditory,chen2026causaltracingaudiotextfusion}.
Interpretability approaches for acoustic models have favored qualitative concept alignment \cite{wu2024andaudionetworkdissection} over quantitative prediction of physical descriptors.
Within the CLAP literature, studies have shown that high-level features such as timbre \cite{deng2025joint} and emotion \cite{velissaridis2025evaluating} are effectively captured. However, low-level acoustic attributes have received far less attention.
\cite{deng2025investigating} shows that applying audio effects at varying intensities traces consistent trajectories in the CLAP embedding space.
The authors exploit this property to improve embedding robustness. 
Conversely, FXEncoder++ \cite{yeh2025fx} learns representations that capture audio effects independently of content.
Text2FX \cite{text2fx} uses CLAP embeddings for audio effect parameter estimation via text prompts, implicitly assuming high cross-modal alignment and encoding of sufficient detail for fine-grained acoustic manipulation.
In contrast, \cite{chung2025listen} proposes an Acoustic Context (ACX) Representation for audio quality enhancement and argues that CLAP embeddings fail to quantify noise or reverberation.

These mixed findings leave a central question open: which low-level acoustic attributes are encoded in CLAP embeddings, and how readily can they be recovered?
Our work aims to address this gap through a systematic probing study.

\begin{figure}
  \centering
  \includegraphics[width=0.48\textwidth]{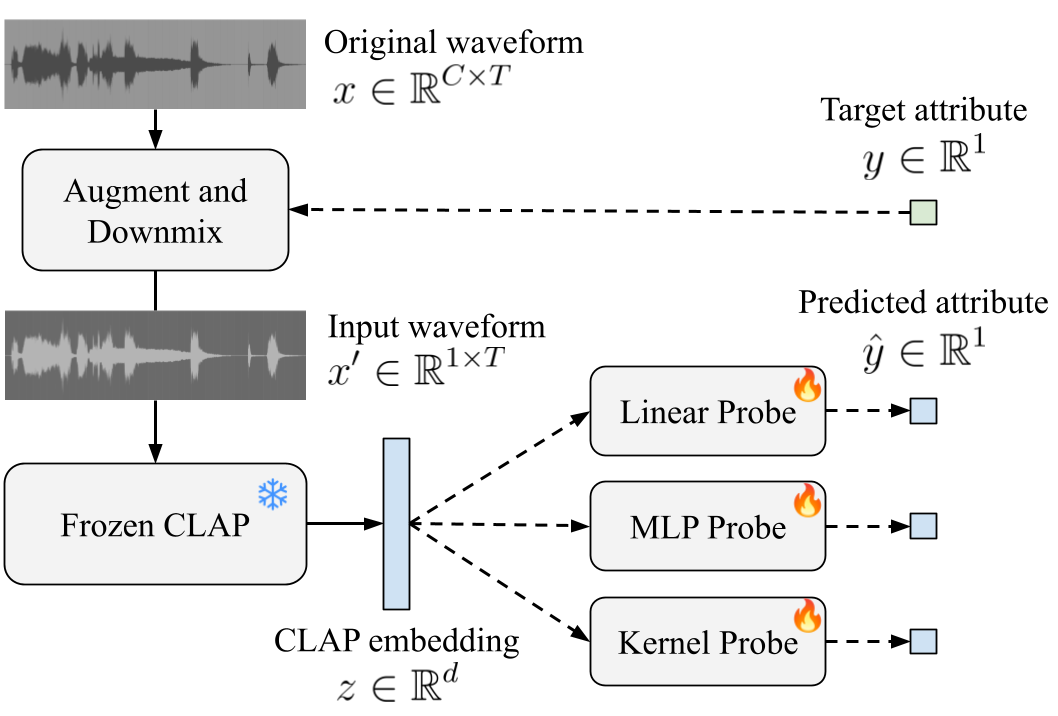}
  \caption{Overview of the probing methodology.}
  \label{fig:method_overview}
\end{figure}  

\section{Methodology}
\label{sec:methodology}

An overview of the probing methodology is presented in \autoref{fig:method_overview}. 
Each waveform $\mathbf{x} \in \mathbb{R}^{C \times T}$, with $C$ channels and $T$ samples, is transformed with an attribute-dependent data augmentation and downmixed to mono obtaining $\mathbf{x^{\prime}} \in \mathbb{R}^{1 \times T}$. $\mathbf{x^{\prime}}$ is then encoded into $\mathbf{z} \in \mathbb{R}^d$ ($d = 512$), which serves as the sole input to all probes. The probes produce the predictions for single scalar attributes $\hat{y} \in \mathbb{R}^1$ and are trained on regression against the ground truth values $y \in \mathbb{R}^1$ used during the augmentation process. The encoder is kept fully frozen: all learned parameters belong exclusively to the probe models.
We use the audio encoder from LAION-CLAP~\cite{laionclap2023,htsatke2022} checkpoint trained on speech, music, AudioSet, and LAION-Audio-630k\footnote{\texttt{laion-clap}: \href{https://github.com/LAION-AI/CLAP}{https://github.com/LAION-AI/CLAP}}, selected for its broad coverage across audio domains.

\subsection{Target Acoustic Attributes} 
\label{sub:target_acoustic_features}

We investigate the encoding of low-level acoustic attributes in the CLAP embedding space, covering three fundamental perceptual dimensions: room acoustics, perceived loudness, and spectral content.
All attributes are defined and computable across all datasets regardless of audio domain or musical instrumentation. %

\textbf{Reverberation Time (RT60).} RT60 measures the time required for sound energy to decay by 60 dB following the cessation of a source, and is one of the most common descriptors of room acoustic conditions~\cite{schroeder1965new}. It is used here as a global measure of the degree of reverberation present in an audio sample, ranging from near-zero values in anechoic or close-miked recordings to several seconds in reverberant spaces.

\textbf{Loudness (LUFS).} Signal loudness is computed as integrated loudness following the ITU-R BS.1770 standard~\cite{itur1770-5}, using the \texttt{pyloudnorm} library\footnote{\texttt{pyloudnorm}: \href{https://github.com/csteinmetz1/pyloudnorm}{https://github.com/csteinmetz1/pyloudnorm}}. Unlike peak amplitude or RMS energy, integrated LUFS accounts for frequency-dependent loudness perception, making it the de facto standard for loudness normalization in broadcast and music production.

\textbf{Spectral Centroid (SC).} The SC is the center of mass of the magnitude spectrum, $\text{SC} = \sum_{k} f_k |X_k| / \sum_{k} |X_k|$, where $k$ denotes the index of the frequency bin, $f_k$ the corresponding frequency, and $X_k$ its STFT magnitude~\cite{muller2015fundamentals}. We compute SC per frame from the STFT magnitude using a Hann window of length 2048 and hop size 512 samples, and then average across frames. We employ the SC rather than the fundamental frequency $f_0$ because it is well-defined for complex audio mixtures such as full songs, for which $f_0$ estimation is unreliable or ill-posed.

\textbf{Relative Pitch (RP).} The RP re-expresses SC on a logarithmic semitone scale as $\text{RP} = 12 \log_2(\text{SC} / f_\text{ref})$, with $f_\text{ref} = 440$\,Hz (A4), following the standard equal-tempered tuning relationship between frequency and pitch~\cite{smith2011spectral, muller2015fundamentals}. RP and SC probe the same underlying spectral property from complementary scales, allowing us to assess whether encoding linearity depends on the scale used.

\subsection{Probe Models} 
\label{sub:probe_models}

We evaluate three probe architectures of increasing complexity. All are intentionally lightweight so that performance differences reflect the embedding space rather than probe capacity.

\textbf{Linear Probe.} A single affine map $\hat{y} = \mathbf{w}^\top \mathbf{z} + b$, with $d{+}1 = 513$ parameters. The weight vector $\mathbf{w}$ defines the \textit{feature axis}: the single direction in embedding space most predictive of the target. Strong performance indicates linear decodability, while poor performance indicates the need for non-linear encoding.

\textbf{MLP Probe.} A two-layer network $\hat{y} = \mathbf{W}_2\, \text{GeLU}(\mathbf{W}_1 \mathbf{z} + \mathbf{b}_1) + b_2$ with $64$ hidden dimensions ($\approx$32.9k parameters). Shallow enough to avoid memorization, it captures non-linear relationships that the linear probe cannot. The MLP--Linear gap quantifies encoding non-linearity.

\textbf{Kernel Probe.} Kernel Ridge Regression (KRR) with an Radial Basis Function (RBF) kernel \cite{scholkopf2002learning} $k(\mathbf{z},\mathbf{z}') = \exp(-\gamma\|\mathbf{z}-\mathbf{z}'\|^2)$, $\gamma = 1.0$. Predictions are given by $\hat{y} = \sum_{j=1}^{M} \alpha_j k(\mathbf{z}, \mathbf{z}_j)$, where $\boldsymbol{\alpha} \in \mathbb{R}^M$ solves the $M{\times}M$ system $(\mathbf{K}+\lambda\mathbf{I})\boldsymbol{\alpha}=\mathbf{y}$, with $\mathbf{y} \in \mathbb{R}^M$ the target values for the $M$ selected examples, and $\lambda = 10^{-3}$. $M$ is capped at $10^4$ randomly selected examples for tractability (KRR scales as $\mathcal{O}(M^3)$). Embeddings are L2-normalized before fitting. Since $\|\hat{\mathbf{z}}-\hat{\mathbf{z}}'\|^2 = 2(1-\cos(\mathbf{z},\mathbf{z}'))$ for unit vectors, the cosine RBF $k(\mathbf{z},\mathbf{z}') = \exp\!\bigl(-2\gamma(1-\cos(\mathbf{z},\mathbf{z}'))\bigr)$ becomes the effective kernel. 
KRR makes no parametric assumptions and serves as a flexible non-linear reference without gradient-based optimisation sensitivity, providing an approximate ceiling for non-linear probe performance.

\subsection{Evaluation Metrics}
\label{sub:evaluation_metrics}

Probes are evaluated using three complementary metrics. 

\textbf{Mean Absolute Error (MAE)} measures prediction error in the original unit of each feature, $\text{MAE} = \frac{1}{N}\sum_{i=1}^{N} |y_i - \hat{y}_i|$. It is directly interpretable but scale-dependent, preventing comparison across features with different units or dynamic ranges.

\textbf{Coefficient of Determination ($R^2$)} is the proportion of target variance explained by the model, $R^2 = 1 - \sum_{i}(y_i - \hat{y}_i)^2 / \sum_{i}(y_i - \bar{y})^2$. $R^2 = 1$ is a perfect fit, $R^2 = 0$ matches a constant mean predictor, and negative values indicate worse-than-trivial performance. Unlike MAE, it is scale-free and accounts for task difficulty via the target variance.

\textbf{Pearson Correlation ($r$)} measures the linear association between predictions and ground truth, $r \in [-1, 1]$. Being insensitive to affine transformations, it is complementary to $R^2$: a high $r$ with low $R^2$ indicates correct direction but miscalibrated scale. Conversely, a negative $r$ indicates an encoding failure, as the probe recovers the attribute in the wrong direction on held-out data.

\section{Experimental settings}
\label{sec:experimental_settings}

\subsection{Datasets}
\label{ssec:datasets}

To test generalization across domains, we evaluate probes on five datasets spanning noise, speech, and music at increasing complexity, from monophonic notes to full music mixtures. All audio is resampled to 48\,kHz to match the sample rate of LAION-CLAP. 

\textbf{White Noise} (100k samples) is a synthetic control of generated white noise samples of 10 seconds each with constant amplitude envelope. Since all samples share the same flat spectral shape and carry no musical or semantic content, all variation in the embeddings is driven exclusively by the applied augmentation, making it the most controlled setting in our evaluation and serving as an upper bound on what a purely acoustic signal can encode. Labels are generated synthetically as described in Sec.~\ref{sub:feature_generation}.

\textbf{NSynth}~\cite{engel2017neural} (306k samples) contains monophonic notes from 1,006 instruments at 16\,kHz, spanning the full MIDI pitch range (21--108) at five velocity levels. Each sample is a 4-second clip in which a single note is held for three seconds and decays over the final second. It represents the simplest scenario in our evaluation, with a single note isolated from rhythmic or harmonic context.

\textbf{VCTK-Corpus}~\cite{Yamagishi2019CSTRVC} (44k samples) is a speech dataset containing recordings from 110 English speakers with diverse accents. The recordings are clean, studio-quality speech with minimal background noise, making them well-suited for isolating acoustic attributes in our study.

\textbf{MusDB18HQ}~\cite{rafii2019musdb18} (34k samples) is a music source separation dataset containing 150 songs at 44.1\,kHz. We use the full additive mix, segmented into 10-second chunks with 1-second overlap. To avoid data leakage, the training, validation, and test sets are split by song. MusDB18HQ represents a more complex musical scenario than NSynth, with multiple instruments, polyphony, and realistic production arrangements. The mixes are raw recordings without compression, equalization, or mastering, making low-level acoustic attributes directly observable.

\textbf{SonicMaster}~\cite{melechovsky2025sonicmaster} (525k samples) comprises songs across 10 balanced genre groups. The dataset contains ground truth mastered tracks and degraded versions with applied audio effects (e.g. equalization, dynamics). We consider only the ground truth mastered tracks and split them by song to ensure no data leakage. This dataset represents the most production-realistic and largest-scale scenario in our benchmark.

For all datasets, targets are measured as per Sec.~\ref{sub:feature_generation} and
we use 80\%/10\%/10\% splits for training, validation, and testing. 
The splits of VCTK and NSynth are speaker-disjoint and instrument-disjoint, respectively.

\subsection{Feature Generation} 
\label{sub:feature_generation}

To ensure that each acoustic attribute spans a balanced distribution of values across its perceptually relevant range, the original datasets are augmented by applying a single targeted perturbation per sample. The embedding and attribute value are then computed from the perturbed signal. Each attribute is augmented independently (i.e., only one perturbation is applied per sample) so that attributes such as RT60 and LUFS cannot co-vary as a result of the augmentation procedure, and any correlation between their learned probe directions reflects embedding structure rather than a data pipeline confound. Following this regime we generate a full copy of each dataset per attribute.

\textbf{RT60:} A target reverberation time is uniformly sampled as $t \sim \mathcal{U}(0.0,\ 2.0)$ s. A synthetic room impulse response (RIR) is generated using \texttt{gpuRIR}\footnote{\texttt{gpuRIR}: \href{https://github.com/DavidDiazGuerra/gpuRIR}{https://github.com/DavidDiazGuerra/gpuRIR}}. 
Room dimensions are sampled uniformly per axis in $[4, 12]$ m, and wall reflection coefficients are derived analytically from $t$ via the Sabine equation \cite{diaz2021gpurir}. The RIR is convolved with the dry waveform, and the output is RMS-normalized to the level of the input to prevent loudness from acting as a proxy cue for RT60.

\textbf{LUFS:} A target loudness level is uniformly sampled as $l \sim \mathcal{U}(-40,\ -10)$ LUFS. The signal is then normalized to $l$ using the loudness normalization function provided by \texttt{pyloudnorm}, ensuring that the training distribution covers a wide and balanced range of perceived loudness levels.

\textbf{SC:} Two procedures are used depending on the dataset. For White Noise, a target SC value is sampled as $f_c \sim \mathcal{U}(500,\ 5000)$ Hz and the noise is filtered with a bandpass filter centered at $f_c$ with a bandwidth drawn from $b \sim \mathcal{U}(0.25,\ 4)$ octaves, controlling the spectral center of mass. 
For all other datasets, a fractional pitch shift $p \sim \mathcal{U}(-6,\ 6)$ semitones is uniformly sampled and applied to the waveform via the \texttt{torch\_audiomentations}\footnote{\texttt{torch\_audiomentations}: \href{https://github.com/iver56/torch-audiomentations}{https://github.com/iver56/torch-audiomentations}} phase vocoder. This creates a smooth distribution of pitch in all datasets, breaking the quantised structures present in music datasets (e.g. NSynth contains only integer pitch values, which we found to hinder regression performance).
The two strategies are not interchangeable by design: bandpass filtering is appropriate only for spectrally flat signals, while pitch shifting preserves content. Since probes are always trained and evaluated within the same dataset, no transfer across augmentation strategies is assumed or performed. In both cases, the SC is computed as defined in Sec.~\ref{sub:target_acoustic_features} using the \texttt{SpectralCentroid} transform from \texttt{torchaudio}\footnote{\texttt{torchaudio}: \href{https://github.com/pytorch/audio}{https://github.com/pytorch/audio}}, yielding a single scalar in Hz.

\textbf{RP:} No additional augmentation is applied for RP. The attribute value is obtained from the SC value following the method described in Sec.~\ref{sub:target_acoustic_features}.

\subsection{Training Details}
\label{sub:training_details}

The Linear and MLP probes are trained for 100 epochs with a batch size of 256 to minimize the MSE objective with AdamW optimizer, using a learning rate of $1 \times 10^{-3}$ and weight decay of $1 \times 10^{-3}$. 
Early stopping is applied based on the validation loss, with a patience of 10 epochs.
The Kernel probe is fitted in a single training iteration. For the largest dataset (SonicMaster), results were verified to be stable when increasing $M$ from $10^4$ to $10^5$, confirming the cap does not constrain performance.
All reported results are averaged over 10 independent runs with different random seeds.

\section{Results}
\label{sec:results}

\setlength{\tabcolsep}{4pt}
\begin{table*}[t]
\caption{Test-set probing results across datasets, averaged over 10 independent seeds. Bold indicates the best or joint-best result per (dataset, feature). %
$\dagger$: probe failure ($R^2 < -1$); negative $R^2$ values above this threshold are reported explicitly.
Standard deviations of $R^2$ across seeds 
are $\le 0.01$ for RT60 and LUFS, $\le 0.05$ for SC (MLP and Kernel only, as Linear probe fails on SC), and $\le 0.04$ for RP.
}
\label{tab:probe_results_all}
\centering
\begin{tabular}{ll|rrr | rrr | rrr | rrr | rrr}
\toprule
 &  & \multicolumn{3}{c|}{White Noise} & \multicolumn{3}{c|}{NSynth} & \multicolumn{3}{c|}{VCTK-Corpus} & \multicolumn{3}{c|}{Musdb18HQ} & \multicolumn{3}{c}{SonicMaster} \\
Feature & Probe & MAE $\downarrow$ & $R^2$ $\uparrow$ & $r$ $\uparrow$ & MAE $\downarrow$ & $R^2$ $\uparrow$ & $r$ $\uparrow$ & MAE $\downarrow$ & $R^2$ $\uparrow$ & $r$ $\uparrow$ & MAE $\downarrow$ & $R^2$ $\uparrow$ & $r$ $\uparrow$ & MAE $\downarrow$ & $R^2$ $\uparrow$ & $r$ $\uparrow$ \\
\midrule

\multirow{3}{*}{RT60} & Linear & 0.09 & 0.95 & 0.98 & 0.25 & 0.67 & 0.82 & 0.12 & 0.92 & 0.96 & 0.21 & 0.77 & 0.88 & 0.21 & 0.76 & 0.87 \\
 & MLP & 0.07 & 0.97 & 0.99 & 0.21 & 0.75 & 0.87 & 0.10 & 0.94 & 0.97 & \textbf{0.20} & \textbf{0.79} & \textbf{0.89} & \textbf{0.19} & \textbf{0.80} & \textbf{0.89} \\
 & Kernel & \textbf{0.04} & \textbf{0.99} & \textbf{1.00} & \textbf{0.18} & \textbf{0.81} & \textbf{0.90} & \textbf{0.09} & \textbf{0.95} & \textbf{0.98} & 0.21 & 0.75 & 0.88 & 0.19 & 0.79 & \textbf{0.89} \\

\midrule

\multirow{3}{*}{LUFS} & Linear & 0.5 & \textbf{0.99} & \textbf{1.00} & 3.0 & 0.80 & 0.90 & 2.1 & 0.92 & 0.96 & 3.5 & 0.76 & 0.88 & 2.1 & 0.90 & 0.95 \\
 & MLP & 0.3 & \textbf{1.00} & \textbf{1.00} & \textbf{2.2} & \textbf{0.89} & \textbf{0.95} & \textbf{1.5} & \textbf{0.95} & \textbf{0.98} & 2.1 & 0.90 & 0.95 & \textbf{1.4} & \textbf{0.96} & \textbf{0.98} \\
 & Kernel & \textbf{0.2} & \textbf{1.00} & \textbf{1.00} & 2.3 & 0.88 & 0.94 & \textbf{1.5} & \textbf{0.95} & \textbf{0.98} & \textbf{1.9} & \textbf{0.92} & \textbf{0.96} & 1.5 & 0.95 & \textbf{0.98} \\

\midrule

\multirow{3}{*}{SC} & Linear & 1451 & $\dagger$ & 0.51 & 183 & 0.43 & 0.68 & 1193 & $\dagger$ & -0.29 & 2930 & $\dagger$ & 0.04 & 729 & -0.06 & 0.35 \\
 & MLP & 119 & 0.98 & 0.99 & \textbf{79} & \textbf{0.90} & \textbf{0.95} & 160 & 0.82 & 0.91 & 574 & 0.40 & 0.67 & \textbf{278} & \textbf{0.84} & \textbf{0.92} \\
 & Kernel & \textbf{34} & \textbf{1.00} & \textbf{1.00} & 94 & 0.85 & 0.93 & \textbf{129} & \textbf{0.89} & \textbf{0.95} & \textbf{467} & \textbf{0.60} & \textbf{0.79} & \textbf{279} & \textbf{0.84} & \textbf{0.92} \\

\midrule

\multirow{3}{*}{RP} & Linear & 1.44 & 0.97 & 0.99 & 4.31 & 0.73 & 0.86 & 2.34 & 0.75 & 0.88 & 3.83 & 0.36 & 0.64 & 2.12 & 0.82 & 0.91 \\
 & MLP & 0.52 & \textbf{1.00} & \textbf{1.00} & \textbf{2.63} & \textbf{0.90} & \textbf{0.95} & \textbf{1.66} & \textbf{0.87} & \textbf{0.94} & 2.78 & 0.64 & 0.81 & \textbf{1.55} & \textbf{0.90} & \textbf{0.95} \\
 & Kernel & \textbf{0.21} & \textbf{1.00} & \textbf{1.00} & 3.22 & 0.85 & 0.92 & 1.69 & \textbf{0.87} & \textbf{0.94} & \textbf{2.63} & \textbf{0.70} & \textbf{0.85} & 1.79 & 0.88 & 0.94 \\

\bottomrule
\end{tabular}
\end{table*}

Results are shown in \autoref{tab:probe_results_all}.
A global trend can be observed: non-linear probes (MLP, Kernel) consistently 
outperform the Linear probe across all features and datasets, with the 
performance margin growing from modest gains for RT60 and LUFS, to substantial 
gains for SC, and moderate gains for RP. This implies that the geometry of 
CLAP's encoding varies considerably across features, which we analyze in more 
detail below.

\textbf{RT60} is strongly and nearly linearly encoded. The Linear probe achieves 
$R^2 \ge 0.67$ and $r \ge 0.82$ across all datasets, and $R^2 \ge 0.92$ on 
White Noise and VCTK-Corpus. MAE ranges from $0.09$\,s on White Noise to 
$0.25$\,s on NSynth. Non-linear probes provide consistent improvements: gains 
are modest on White Noise and VCTK-Corpus ($\Delta R^2 \le 0.04$) but more 
substantial on NSynth, where the Kernel probe reaches $R^2 = 0.81$ vs.\ the 
Linear probe's $R^2 = 0.67$. The Kernel probe reaches $R^2 \ge 0.75$ across 
all datasets ($R^2 = 0.99$ on White Noise), confirming a predominantly linear 
underlying structure.

\textbf{LUFS} is also strongly and approximately linearly encoded. The Linear 
probe achieves $R^2 \ge 0.76$ and $r \ge 0.88$ (MAE: $2.1$--$3.5$\,LUFS 
across real datasets). Non-linear probes bring moderate improvements 
($R^2 \ge 0.88$, $r \ge 0.94$, MAE: $1.4$--$2.3$\,LUFS), with MLP and 
Kernel performing comparably across most datasets.

\textbf{SC} results substantially differ from RT60 and LUFS. 
The Linear probe performs at or below chance on four of five datasets.
On VCTK-Corpus and MusDB18HQ ($\dagger$) with $r = -0.29$ and 
$r = 0.04$ respectively, on SonicMaster $R^2 = -0.06$ with $r = 0.35$, and 
on White Noise ($\dagger$) with $r = 0.51$, indicating the correct 
direction was partially identified but at a severely miscalibrated scale. 
NSynth is a notable exception: the Linear probe achieves $R^2 = 0.43$ 
($r = 0.68$), a partial linear recovery not observed on any other dataset, 
consistent with its simplified acoustic content (isolated monophonic notes) 
producing a particularly clean SC signal in the embedding space. Non-linear 
probes recover SC substantially better across all datasets: the MLP reaches 
$R^2 = 0.84$ on SonicMaster, and the Kernel probe achieves the best results 
overall ($R^2 \ge 0.60$, $r \ge 0.79$, MAE $\le 467$\,Hz). The predominant 
failure of the Linear probe and consistent success of non-linear probes 
evidences that SC generally lies on a curved manifold that a single linear 
projection cannot recover, with NSynth as a boundary case under simplified 
acoustic conditions.

\textbf{RP} is linearly encoded across most datasets, though with varying 
strength. The Linear probe achieves strong performance on White Noise 
($R^2 = 0.97$), SonicMaster ($R^2 = 0.82$, $r = 0.91$), and VCTK-Corpus 
($R^2 = 0.75$, $r = 0.88$), moderate performance on NSynth ($R^2 = 0.73$, 
$r = 0.86$), and substantially weaker performance on MusDB18HQ ($R^2 = 0.36$, 
$r = 0.64$). The weaker MusDB18HQ result is consistent with both its smaller 
training set size and the greater embedding variation introduced by full 
polyphonic mixes, which reduces the proportion of variance attributable to a 
single linear direction. In all cases, $r > 0$ confirms that the linear probe 
always identifies the correct direction in embedding space. This stands in 
sharp contrast to SC, a monotone transformation of RP yet one for which the 
Linear probe is essentially uninformative on most datasets: SC yields 
$r = -0.29$ on VCTK-Corpus and $r = 0.04$ on MusDB18HQ, with the negative 
value on VCTK-Corpus indicating that the Linear probe recovers SC in the wrong 
direction entirely. Non-linear probes recover RP reliably across all datasets 
($R^2 \ge 0.64$, $r \ge 0.81$, with $R^2 = 1.00$ on White Noise), confirming 
that RP information is present regardless of poor linear decodability.
Two factors may explain the SC/RP divergence: the log transform produces a more uniform target distribution that is intrinsically easier to fit linearly, 
and because CLAP operates on log-mel spectrograms, pitch shifts may be preserved as a near-linear direction in embedding space.

\label{sub:ablation_study}

\subsection{Feature Axis Analysis}
\label{sub:ablation_relevance_axis}

All figures and numerical results correspond to the training run with the median test $R^2$ across the 10 independently seeded runs.

To quantify the geometric alignment of learned feature directions across datasets, we compute the pairwise cosine similarity $\cos(\mathbf{w}_i, \mathbf{w}_j) = \mathbf{w}_i^\top \mathbf{w}_j / (\|\mathbf{w}_i\| \cdot \|\mathbf{w}_j\|)$ between each pair of weight vectors trained independently on datasets $i$, $j$. The corresponding angle is $\theta = \arccos(|\cos(\mathbf{w}_i, \mathbf{w}_j)|) \in [0^\circ, 90^\circ]$. The absolute value accounts for the sign indeterminacy of the linear probe weight vector: $\mathbf{w}$ and $-\mathbf{w}$ define the same feature axis.
For reference, in $d = 512$ dimensions, two random unit vectors have expected absolute cosine similarity $\mathbb{E}[|\cos|] \approx \sqrt{2/(\pi d)} \approx 0.035$ ($\theta \approx 88^\circ$) for large $d$, i.e.,\ nearly orthogonal by chance.

\begin{figure}[h]
  \centering
  \includegraphics[width=\linewidth]{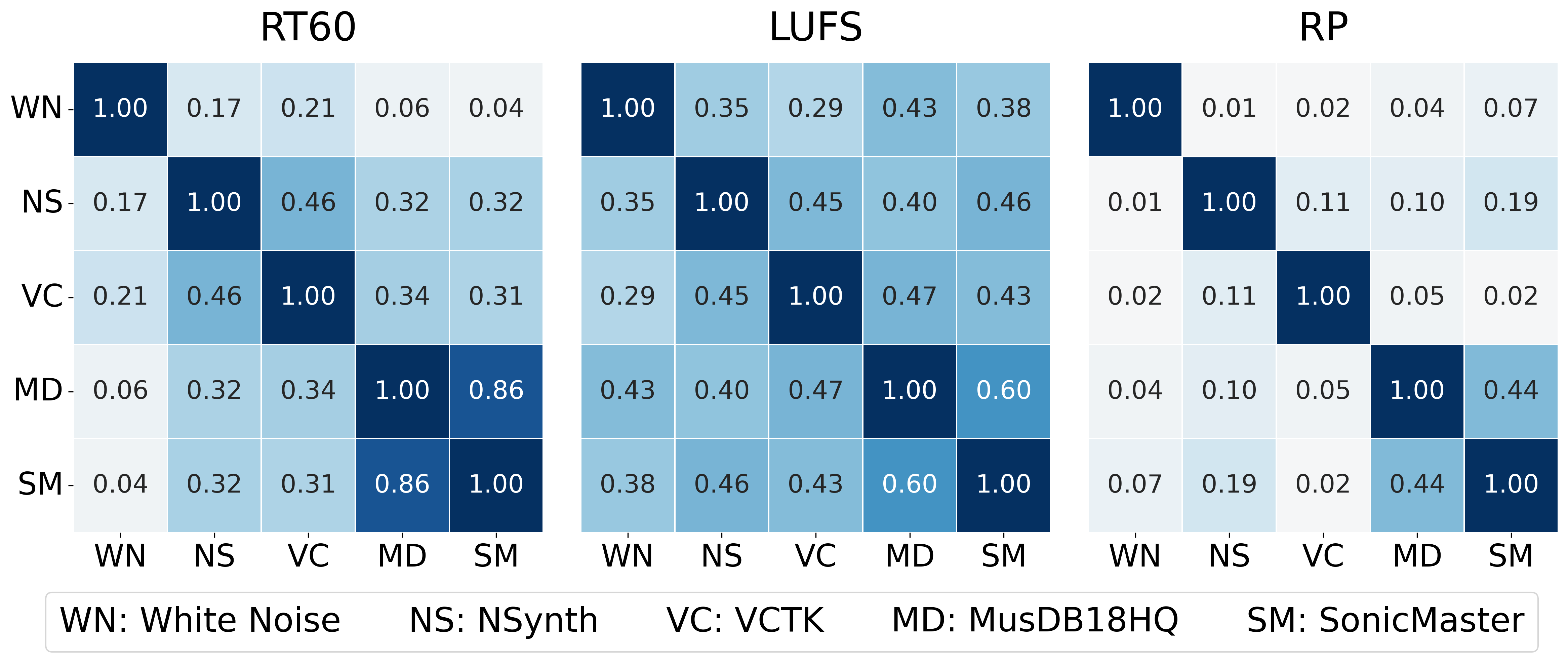}
  \caption{Pairwise cosine similarity of linear probe weight vectors $\mathbf{w} \in \mathbb{R}^{512}$ trained independently on each dataset, for RT60 (left), LUFS (center), and RP (right). 
  }
  \label{fig:axis_cosine_similarity}
\end{figure}

As shown in \autoref{fig:axis_cosine_similarity}, RT60 and LUFS feature axes are consistent across datasets, while the RP axis is highly domain-dependent.

\begin{figure*}[ht]
    \centering
    \begin{subfigure}[b]{0.33\linewidth}
        \includegraphics[width=\linewidth]{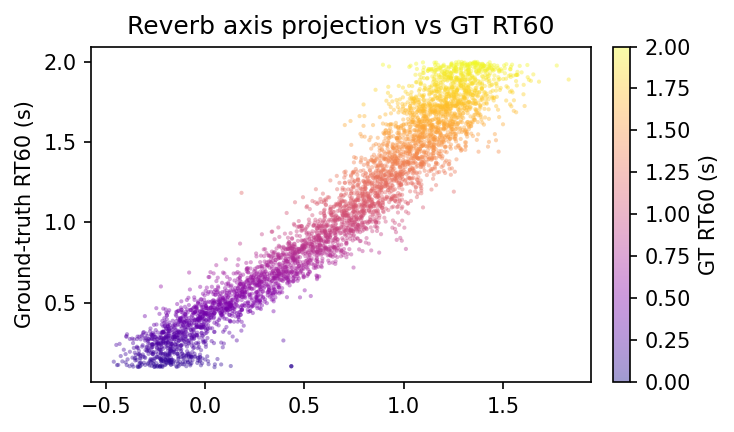}
        \caption{RT60 axis projection vs.\ ground truth.}
        \label{fig:rt60_axis_scatter}
    \end{subfigure}
    \hfill
    \begin{subfigure}[b]{0.33\linewidth}
        \includegraphics[width=\linewidth]{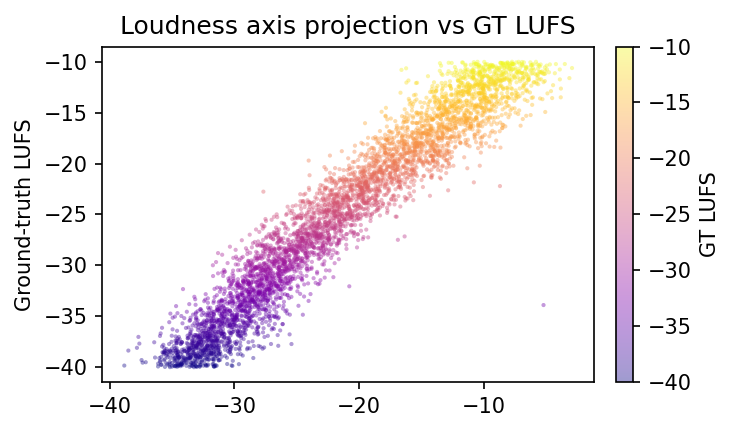}
        \caption{LUFS axis projection vs.\ ground truth.}
        \label{fig:lufs_axis_scatter}
    \end{subfigure}
    \hfill
    \begin{subfigure}[b]{0.33\linewidth}
        \includegraphics[width=\linewidth]{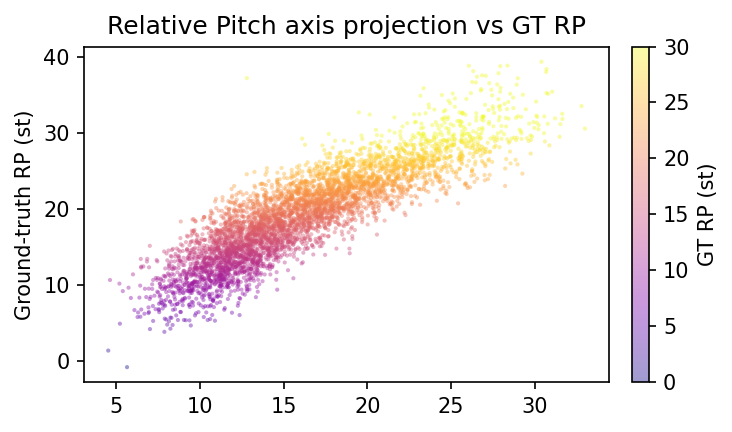}
        \caption{RP axis projection vs.\ ground truth.}
        \label{fig:pitch_axis_scatter}
    \end{subfigure}
    \caption{VCTK test-set embeddings projected onto the linear-probe weight vector ($x$-axis) vs.\ ground-truth value (GT, $y$-axis), colored by target value. Both RT60 and LUFS show a monotone band with mild curvature, confirming a predominantly linear structure consistent with $R^2 = 0.92$ for the Linear probe on VCTK-Corpus (\autoref{tab:probe_results_all}). RP shows a more pronounced curve, consistent with a weaker linear encoding ($R^2 = 0.75$). SC is omitted as the Linear probe fails on VCTK-Corpus ($\dagger$), making the projection uninformative.}
    \label{fig:axis_projections}
\end{figure*}

\textbf{RT60} axes show moderate cross-domain alignment ($\cos \in [0.17,\, 0.46], \theta \in [62^\circ,\, 80^\circ]$). 
MusDB18HQ--SonicMaster is the strongest pair ($\cos = 0.86$, $\theta \approx 31^\circ$), attributable to their shared music-mixture domain. 
NSynth and VCTK-Corpus align more strongly with each other ($\cos = 0.46, \theta \approx 62^\circ$) than either does with the complex music mixtures ($\cos \in [0.31,\, 0.34]$, and $\theta \in [70^\circ,\, 72^\circ]$). Both are single-source datasets despite belonging to different domains, suggesting that content complexity is a strong organizing principle. 
Pairs involving White Noise fall at the lower end ($\cos \in [0.04,\, 0.21], \theta \in [78^\circ,\, 88^\circ]$), producing a probe direction with no content-driven components. This is consistent with the absence of salient content in the White Noise dataset.
Excluding White Noise, all pairs substantially exceed the chance baseline. This is consistent with, though not definitive evidence of, a common RT60 direction across domains. %

\textbf{LUFS} axes are uniformly consistent ($\cos \in [0.29,\, 0.60]$, $\theta \in [53^\circ,\, 73^\circ]$), including pairs involving White Noise. The higher cross-domain consistency relative to RT60 is consistent with gain scaling being a globally uniform transformation that dominates the embedding displacement regardless of content. Residual variation nonetheless indicates that domain-specific components remain present in all estimated axes.

\textbf{RP} presents a markedly different picture: with the exception of MusDB18HQ--SonicMaster ($\cos = 0.44$, $\theta = 64^\circ$), while most pairs are near or at the chance baseline, indicating near-orthogonal axes across domains. The one coherent pair shares a common music-mixture domain, suggesting that a common RP direction is recoverable only within a domain. Unlike RT60 and LUFS, whose global transformations leave a partially consistent trace across domains, the RP axis is dominated by domain-specific embedding structure and does not transfer across speech, isolated notes, noise, and music, even though RP is well-encoded and linearly recoverable within each domain individually.

\autoref{fig:axis_projections} illustrates what projecting onto a linear probe weight vector looks like in practice, using VCTK-Corpus as a representative dataset. Each embedding of the test set is projected onto the scalar $p = \mathbf{w}^\top \mathbf{z} / \|\mathbf{w}\|$ and plotted against its ground-truth value.

\subsection{Effect of Training Set Size} 
\label{sub:ablation_dataset_size}

To understand how probe performance scales with the amount of training data, we train Linear probes on progressively smaller subsets of SonicMaster and measure performance as a function of training set size. SonicMaster is chosen for this study as it is the largest and most domain-diverse dataset in our benchmark, making it a suitable basis for a data efficiency analysis. SC is excluded since the Linear probe fails at all training sizes. The Linear probe is used here by design: with only $d{+}1$ parameters, its saturation curve directly reflects the sample complexity of estimating a single direction in the embedding space, without confounds from probe capacity or gradient-based optimisation sensitivity that would arise with the MLP or Kernel probes.

\begin{table}[h]
  \centering
  \caption{Linear probe test-set MAE / $R^2$ on SonicMaster at varying training-set sizes (val/test splits always use the full set). $\dagger$: probe failure ($R^2 < -1$); negative $R^2$ values above this threshold are reported explicitly. SC is omitted as the Linear probe fails at all sizes. Pearson $r$ is omitted from this and the following table as it is monotonically related to $R^2$ on a fixed test set.}
  \label{tab:size_ablation_linear}
  \setlength{\tabcolsep}{3.5pt}
  \footnotesize
  \begin{tabular}{r r rr rr rr}
    \toprule
    & & \multicolumn{2}{c}{RT60} & \multicolumn{2}{c}{LUFS}
    & \multicolumn{2}{c}{RP} \\
    \cmidrule(lr){3-4}\cmidrule(lr){5-6}\cmidrule(lr){7-8}
    Ratio & $N\,(\times 10^3)$ & MAE & $R^2$ & MAE & $R^2$ & MAE & $R^2$ \\
    \midrule
    0.01  & 4.2   & 0.22 & 0.73 & 17.6 & $\dagger$ & 22.6 & $\dagger$ \\
    0.02  & 8.4   & 0.22 & 0.75 &  5.7 &  0.34 &  6.5 & $-$0.71 \\
    0.05  & 21.0  & \textbf{0.21} & \textbf{0.76} &  3.7 &  0.72 &  3.4 &    0.51 \\
    0.10  & 42.1  & \textbf{0.21} & \textbf{0.76} &  2.7 &  0.85 &  2.6 &    0.71 \\
    0.20  & 84.1  & \textbf{0.21} & \textbf{0.76} & \textbf{2.3} &  0.88 & \textbf{2.2} &    0.78 \\
    0.50  & 210.0 & \textbf{0.21} & \textbf{0.76} & \textbf{2.3} & \textbf{0.89} & \textbf{2.2} & \textbf{0.79} \\
    1.00  & 421.0 & \textbf{0.21} & \textbf{0.76} & \textbf{2.3} & \textbf{0.89} & \textbf{2.2} & \textbf{0.79} \\
    \bottomrule
  \end{tabular}
\end{table}

\autoref{tab:size_ablation_linear} shows markedly different data efficiency profiles across features. 
RT60 is robust to limited data: the Linear probe achieves $R^2 = 0.73$ with as few as 4.2k training examples (ratio 0.01), and performance saturates at $R^2 = 0.76$ from ratio 0.05 onward. This suggests that the RT60 direction in the embedding space is high-variance and salient. 
LUFS and RP are considerably more data-hungry. LUFS improves monotonically from complete failure at ratio 0.01 to saturation at ratio 0.50 ($R^2 = 0.89$). RP remains negative at ratio 0.02 ($R^2 = -0.71$) before recovering from ratio 0.05 onward and saturating at ratio 0.50 ($R^2 = 0.79$). At least 21k examples (ratio 0.05) are needed before the probe first yields positive $R^2$, and saturation is reached at 210k examples (ratio 0.50).

\subsection{Generalization to Other Models}
\label{sub:ablation_other_models}

To assess whether the encoding patterns observed in CLAP are specific to this model or arise more broadly from large-scale audio pretraining, we repeat the linear probe evaluation on the following additional audio embedding models: 
MS-CLAP versions of 2022~\cite{elizalde2022clap} and 2023~\cite{elizalde2023clap}, 
MERT~\cite{li2023mert}, 
Wav2Vec2~\cite{baevski2020wav2vec},
WavLM (Base and Large)~\cite{chen2022wavlm},
Whisper~\cite{radford2023robust},
and VGGish~\cite{hershey2017cnn}. 
As most of these models are trained and used in downstream tasks in the speech domain, the natural choice for this study is VCTK-Corpus.

Embeddings are extracted via \texttt{fadtk}~\cite{fadtk}\footnote{\texttt{fadtk}: \href{https://github.com/microsoft/fadtk}{https://github.com/microsoft/fadtk}} using the same training and evaluation protocol as Sec.~\ref{sub:probe_models}, with probe dimensionality adapted to each model's embedding size $d$ (MLP hidden size fixed at 64), and audio resampled to each model's sample rate.

\begin{table}[h]
  \centering
  \caption{Multi-embedder probe evaluation on VCTK-Corpus. MAE $\downarrow$, $R^2$ $\uparrow$; best or joint-best probe per (embedder, feature) in \textbf{bold}. $\dagger$: probe failure ($R^2 < -1$); negative $R^2$ values above this threshold are reported explicitly.}
  \label{tab:multi_embedder_vctk}
  \setlength{\tabcolsep}{2.5pt}
  \footnotesize
  \begin{tabular}{ll rr rr rr rr}
    \toprule
    & & \multicolumn{2}{c}{\textbf{RT60}} & \multicolumn{2}{c}{\textbf{LUFS}} & \multicolumn{2}{c}{\textbf{SC}} & \multicolumn{2}{c}{\textbf{RP}} \\
    \cmidrule(lr){3-4}\cmidrule(lr){5-6}\cmidrule(lr){7-8}\cmidrule(lr){9-10}
    Embedder & Probe & MAE & $R^2$ & MAE & $R^2$ & MAE & $R^2$ & MAE & $R^2$ \\
    \midrule
    \multirow{3}{*}{\shortstack[l]{MS-CLAP\\2022 (1024)}} & Linear & 0.14 & 0.89 & 4.6 & 0.54 & 233 & 0.60 & \textbf{2.2} & \textbf{0.78} \\
    & MLP & 0.13 & 0.90 & \textbf{4.2} & \textbf{0.59} & 168 & 0.82 & \textbf{2.2} & \textbf{0.78} \\
    & Kernel & \textbf{0.12} & \textbf{0.92} & 4.3 & 0.58 & \textbf{167} & \textbf{0.82} & \textbf{2.2} & \textbf{0.78} \\
    \midrule
    \multirow{3}{*}{\shortstack[l]{MS-CLAP\\2023 (1024)}} & Linear & 0.08 & 0.96 & 1.7 & 0.93 & 213 & 0.67 & \textbf{1.5} & 0.89 \\
     & MLP & \textbf{0.07} & \textbf{0.97} & \textbf{1.6} & \textbf{0.94} & 121 & 0.90 & \textbf{1.5} & \textbf{0.90} \\
     & Kernel & \textbf{0.07} & \textbf{0.97} & 1.7 & 0.93 & \textbf{118} & \textbf{0.91} & \textbf{1.5} & 0.89 \\
    \midrule
    \multirow{3}{*}{\shortstack[l]{MERT\\(768)}} & Linear & 0.10 & 0.94 & \textbf{7.4} & -0.02 & 494 & -0.79 & 1.6 & 0.88 \\
     & MLP & 0.10 & 0.95 & \textbf{7.4} & \textbf{-0.01} & 119 & 0.91 & 1.3 & 0.92 \\
     & Kernel & \textbf{0.08} & \textbf{0.96} & 8.2 & -0.29 & \textbf{93} & \textbf{0.94} & \textbf{1.2} & \textbf{0.93} \\
    \midrule
    \multirow{3}{*}{\shortstack[l]{Wav2Vec2\\(768)}} & Linear & 0.11 & 0.93 & \textbf{7.5} & \textbf{0.00} & 373 & -0.03 & 2.2 & 0.79 \\
     & MLP & 0.11 & 0.93 & \textbf{7.5} & \textbf{0.00} & 160 & 0.84 & \textbf{1.7} & 0.86 \\
     & Kernel & \textbf{0.09} & \textbf{0.95} & 8.8 & -0.50 & \textbf{126} & \textbf{0.90} & \textbf{1.7} & \textbf{0.87} \\
    \midrule
    \multirow{3}{*}{\shortstack[l]{WavLM-B\\ (768)}} & Linear & 0.12 & 0.91 & 3.8 & 0.68 & 496 & -0.82 & 1.8 & 0.84 \\
     & MLP & 0.11 & 0.93 & \textbf{3.0} & \textbf{0.81} & 136 & 0.88 & \textbf{1.4} & \textbf{0.90} \\
     & Kernel & \textbf{0.10} & \textbf{0.94} & 3.1 & 0.79 & \textbf{108} & \textbf{0.93} & \textbf{1.4} & \textbf{0.90} \\
    \midrule
    \multirow{3}{*}{\shortstack[l]{WavLM-L\\(1024)}} & Linear & 0.07 & 0.97 & \textbf{7.5} & \textbf{-0.01} & 490 & -0.87 & 1.7 & 0.86 \\
    & MLP & 0.07 & 0.97 & \textbf{7.5} & \textbf{-0.01} & 119 & 0.91 & \textbf{1.4} & \textbf{0.91} \\
    & Kernel & \textbf{0.05} & \textbf{0.98} & 8.3 & -0.31 & \textbf{103} & \textbf{0.93} & \textbf{1.4} & \textbf{0.91} \\
    \midrule
    \multirow{3}{*}{\shortstack[l]{Whisper\\(512)}} & Linear & 0.10 & 0.94 & 1.1 & 0.97 & 426 & -0.31 & 2.0 & 0.82 \\
     & MLP & 0.10 & 0.95 & 0.8 & 0.98 & 154 & 0.85 & \textbf{1.5} & \textbf{0.89} \\
     & Kernel & \textbf{0.09} & \textbf{0.96} & \textbf{0.8} & \textbf{0.99} & \textbf{119} & \textbf{0.91} & \textbf{1.5} & \textbf{0.89} \\
    \midrule
    \multirow{3}{*}{\shortstack[l]{VGGish\\(128)}} & Linear & 0.15 & 0.87 & 4.0 & 0.68 & 1130 & $\dagger$ & 2.4 & 0.74 \\
     & MLP & 0.11 & 0.93 & 2.6 & 0.86 & 168 & 0.83 & \textbf{1.9} & \textbf{0.85} \\
     & Kernel & \textbf{0.09} & \textbf{0.95} & \textbf{2.2} & \textbf{0.89} & \textbf{146} & \textbf{0.86} & \textbf{1.9} & \textbf{0.85} \\
    \bottomrule
  \end{tabular}
\end{table}

\autoref{tab:multi_embedder_vctk} shows that the global trend from Sec.~\ref{sec:results} holds across all models for RT60, SC, and RP. LUFS, however, varies substantially by architecture.

\textbf{RT60} is strongly and linearly encoded across all models (Linear probe $R^2 \ge 0.87$), which confirms that this trend generalizes beyond LAION-CLAP.

\textbf{LUFS} is well encoded in Whisper ($R^2 = 0.97$), which yields the best linear result across all models, and MS-CLAP 2023 ($R^2 = 0.93$). WavLM-Base and VGGish show weaker linear encoding ($R^2 = 0.68$), and non-linear probes improve performance ($R^2 = 0.81$ and $R^2 = 0.89$, respectively). MS-CLAP 2022 shows substantially weaker LUFS encoding ($R^2 = 0.54$ linear, $0.59$ best probe) despite sharing the same contrastive audio-text objective as the 2023 version. Given that the two checkpoints differ in both backbone architecture (CNN14 vs.\ HTSAT-tiny) and training data scope, the precise cause of this difference cannot be isolated from the available evidence. 
All probes yield $R^2 \approx 0$ or below on MERT, Wav2Vec2 and WavLM-Large, indicating that loudness is absent from those representations rather than merely non-linearly encoded. 
In all cases, the failure is explained by architectural amplitude invariance: Wav2Vec2 and WavLM-Large normalize inputs to zero mean and unit variance before encoding, while MERT applies it in its convolutional feature extractor, which removes global amplitude scaling from all downstream activations. %

\textbf{SC} remains non-linearly encoded across all non MS-CLAP models ($R^2 \le 0$). 
Both MS-CLAP versions achieve partial linear recovery ($R^2 = 0.67$ and $0.60$ for the 2023 and 2022 checkpoints respectively). 
This result is remarkable: the two checkpoints differ in both audio encoder architecture (CNN14 vs.\ HTSAT-tiny) and training data scope, yet both exhibit partial linear SC recovery that no other model replicates.

\textbf{RP} is strongly and linearly encoded across all models ($R^2 \ge 0.74$). Non-linear probes bring modest additional gains for most models (Kernel $R^2$ ranging from $0.85$ for VGGish to $0.93$ for MERT). 
An exception is MS-CLAP 2022, where all three probes saturate at $R^2 = 0.78$, indicating that the embedding contains no additional RP information beyond what is linearly encoded.

\subsection{Text-based Attribute Prediction} 
\label{sub:text_prediction}

CLAP's joint embedding space enables zero-shot acoustic attribute prediction: the audio-trained linear probe applied directly to text embeddings converts a natural language description into a predicted attribute value. We present this as a qualitative demonstration of cross-modal consistency, as the reduced and hand-crafted prompt set is not designed for statistical generalization.

\begin{figure}[h] 
    \centering 
    \includegraphics[width=\linewidth]{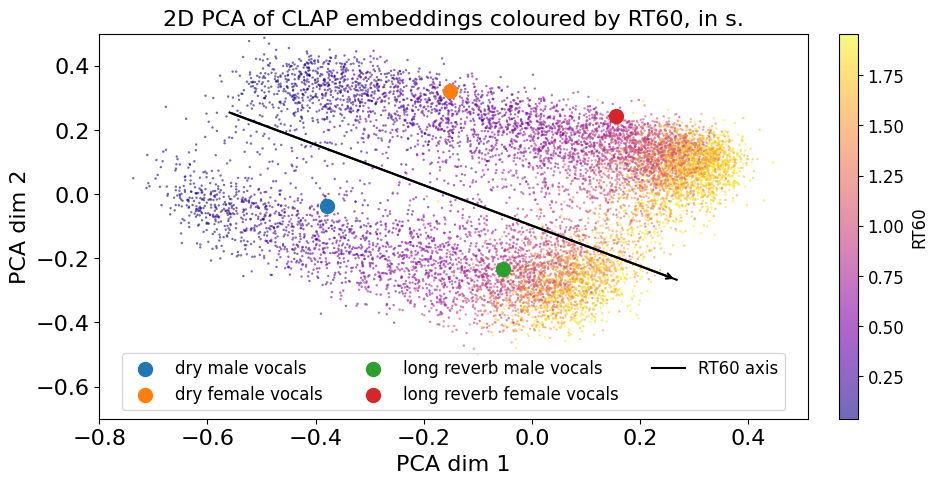} 
    \caption{PCA projection of LAION-CLAP embeddings for the VCTK-Corpus test set, colored by RT60. Projected text descriptions confirm that the qualitative descriptors \textit{dry} and \textit{long reverb} land at geometrically coherent positions along the RT60 axis identified by the Linear probe (black arrow).} \label{fig:vctk_rt60_pca} 
\end{figure}

\autoref{fig:vctk_rt60_pca} shows the PCA projection of VCTK-Corpus test-set embeddings, colored by RT60. A monotonically increasing RT60 direction runs from upper left to lower right, consistent with the Linear probe axis (black arrow). Two illustrative descriptors, \textit{dry} and \textit{long reverb}, land at coherent positions along this axis.

\begin{table}[h]
\centering
\caption{Predicted RT60 values obtained by applying the VCTK-Corpus linear probe to CLAP text embeddings. Acoustic descriptors (above rule) and semantic controls (below rule) are ordered by expected RT60.}
\label{tab:rt60_text}
\footnotesize
\begin{tabular}{lc}
\toprule
Description & Predicted RT60 (s) \\
\midrule
anechoic               & \phantom{$-$}0.08 \\
dry                    & $-$0.08           \\
slightly reverberant   & \phantom{$-$}0.82 \\
moderately reverberant & \phantom{$-$}1.10 \\
highly reverberant     & \phantom{$-$}1.80 \\
\midrule
recording studio       & \phantom{$-$}0.09 \\
living room            & \phantom{$-$}0.29 \\
concert hall           & \phantom{$-$}1.13 \\
cathedral              & \phantom{$-$}0.89 \\
\bottomrule
\end{tabular}
\end{table}

The five descriptions in \autoref{tab:rt60_text} form a monotonic sequence from near-zero to 1.80\,s. The negative prediction for \textit{dry} ($-$0.08\,s) is an artefact of the unconstrained probe near the lower training boundary ($\mathcal{U}(0.0,\ 2.0)$\,s), and its absolute value (0.08\,s) is coherent with \textit{anechoic}. \textit{Recording studio} (0.09\,s), \textit{living room} (0.29\,s), and \textit{concert hall} (1.13\,s) form a monotonically increasing sequence, while \textit{cathedral} yields only 0.89\,s, well below the physically expected $>$2\,s, which lies outside the training range $\mathcal{U}(0.0,\ 2.0)$\,s. 

Repeating the same experiment with LUFS produced no consistent ordering, contrasting with the strong audio-side encoding (Linear $R^2 = 0.92$ on VCTK-Corpus). In addition, a similar demonstration for RP is not straightforward: pitch descriptors in natural language are inherently relational, lacking a fixed absolute reference analogous to reverberation decay time, and the domain-dependence of the RP axis identified in Sec.~\ref{sub:ablation_relevance_axis} further limits the interpretability of any text-side prediction. 
Improving audio--text alignment for attributes with relational or domain-dependent language representations remains an open direction for future work.

\section{Conclusion}
\label{sec:conclusion}

We presented a systematic probing study of three fundamental perceptual dimensions, reverberation (RT60), loudness (LUFS), and spectral content (SC and RP), in CLAP audio embeddings, using Linear, MLP, and Kernel Ridge Regression probes across five datasets spanning noise, speech, and music.

Our primary finding is that all the attributes we probed are well represented in CLAP and other audio foundation model embeddings: non-linear probes recover each attribute reliably across all datasets and model families. Within this broader picture, two encoding regimes emerge: RT60, LUFS, and RP are approximately linearly encoded and recoverable with a simple linear probe, while SC requires non-linear probes. 
For the linearly encoded features, the identified feature-axis directions are geometrically consistent for RT60 and LUFS across datasets, reflecting intrinsic embedding structure. For RP, in contrast, the feature-axis directions have been found to be domain-specific. 
Both regimes generalize across eight additional audio foundation models, with the exception that amplitude-invariant architectures (Wav2Vec2, WavLM-Large, and MERT) discard loudness entirely by construction. 
Notably, both MS-CLAP versions and LAION-CLAP are the only models achieving partial linear SC recovery in at least one domain, suggesting that spectral content may become linearly encoded under certain architectural, training conditions, or data regimes.

These findings have direct implications for audio practitioners: the encoding of all three perceptual dimensions in a shared embedding space opens the possibility of exploiting a single frozen foundation model for simultaneous estimation of reverberation, loudness, and spectral content. The geometric consistency of RT60 and LUFS feature axes further hints that such directions could inform workflows such as automatic mix analysis, effect-chain estimation, and text-driven effect control, though validating these applications remains future work. Limitations include the exclusive focus on final-layer embeddings, the use of shoebox room geometry for RT60 augmentation, and residual reverberation in music mixtures (MusDB18HQ, SonicMaster): although the applied RIR dominates the target RT60, these mixes cannot be guaranteed fully dry, so a small non-zero RT60 may be present prior to augmentation. Extending the attribute set (to e.g. tempo, dynamics, and stereo imaging) is a natural next step.

\bibliographystyle{IEEEtranDAFx}
\bibliography{references} %

\end{document}